# Finding quasinormal modes directly from the boundary conditions in a Schwarzschild black hole

Jeff Steinhauer

*Department of Physics, Technion – Israel Institute of Technology; Haifa, 3200003, Israel*

We present a conceptually simple method for finding quasinormal modes (QNMs) in a Schwarzschild black hole. QNMs are defined by their boundary conditions at infinity and at the horizon. These include the intuitively satisfying assertion that the QNM should be purely outgoing at infinity, with no incoming wave coming from infinity and scattering off the black hole. One applies this condition at past null infinity by demanding that the incoming wave there vanish, despite the infinite outgoing wave at the same point in spacetime. We search for QNMs by minimizing the incoming wave, but we are forced to work at a finite distance rather than past null infinity, to avoid the infinite outgoing wave. This technique lets us find the fundamental QNM, but we do not succeed in finding the overtones since the incoming waves always vanish due to damping. The technique is inherently approximate due to the boundary condition at a finite distance.

Press found that a perturbed Schwarzschild black hole exhibits a damped vibration with a characteristic frequency ~ $l/\sqrt{27}$, where $l$ is the spherical-harmonic index, and he dubbed this oscillation a quasinormal mode (QNM) [1]. This same QNM frequency appeared for a Schwarzschild black hole perturbed by an infalling particle [2], and during black hole formation via collapse [3]. QNMs are defined by the condition of purely outgoing waves at infinity and purely ingoing waves at the horizon. Chandrasekhar constructed a series solution to the wave equation which was guaranteed to meet these boundary conditions for any frequency. A discrete spectrum resulted from the requirement of smoothness at intermediate distances [4]. Leaver found the QNM frequencies while completely avoiding computing the wavefunctions [5]. This work is an attempt to find QNM frequencies directly from the ingoing/outgoing boundary conditions, although we apply them at intermediate distances. For a given frequency, we find a solution to the wave equation which meets the incoming boundary condition, and evaluate the extent to which the solution meets the outgoing boundary condition. This procedure allows us to find the fundamental QNMs, but not the overtones due to their high damping.

As a first step, we study the meaning of the boundary conditions via the master function $\Psi(t, r_*) = \psi(r_*)e^{i\omega t}$ in the limit $r_* \to \pm\infty$ ($r_*$ is the tortoise coordinate), where the top and bottom signs correspond to the boundaries at infinity and at the event horizon, respectively. The complex frequency is given by $\omega \equiv \omega_\mathrm{R} + i\omega_\mathrm{I}$. We are interested in $\omega_\mathrm{I} > 0$ so that the mode decays with $t$ [4]. The function $\psi$ is the solution of the wave equation [6, 7]

$$-\frac{\partial^2\psi}{\partial {r_*}^2} + V(r_*)\psi = \omega^2\psi \tag{1}$$

where $V$ is a finite localized potential. In regions of $r_*$ with $V \approx 0$ (i.e., far from the maximum in the effective potential), the general solution of Eq. 1 is

$$\psi \approx A_\mathrm{out}e^{-i\omega r_*} + A_\mathrm{in}e^{i\omega r_*} \tag{2}$$

where the first term is outgoing and the second term is incoming. Writing $\omega$ in terms of its components,

$$\psi \approx A_{\text{out}} e^{\omega_{\text{I}} r_*} e^{-i\omega_{\text{R}} r_*} + A_{\text{in}} e^{-\omega_{\text{I}} r_*} e^{i\omega_{\text{R}} r_*} \tag{3}$$

For $r_* \to +\infty$, the outgoing term is infinitely larger than the incoming term for any frequency with $\omega_{\text{I}} > 0$, as long as $A_{\text{out}}$ and $A_{\text{in}}$ are finite, due to the factors $e^{\pm\omega_{\text{I}} r_*}$. Similarly, the incoming term is infinitely larger for $r_* \to -\infty$, for any frequency (with $\omega_{\text{I}} > 0$). This holds on all the boundaries of the domain of outer communications in the Schwarzschild spacetime shown in Fig. 1, including future null infinity $\mathcal{I}^+$, past null infinity $\mathcal{I}^-$, spatial infinity $i^0$, the future event horizon $\mathcal{H}^+$, the past event horizon $\mathcal{H}^-$, and the bifurcation point $\mathcal{B}$. We can further emphasize this by multiplying Eq. 2 by the time dependence $\exp(i\omega t)$, which gives

$$\Psi(t, r_*) \approx A_{\text{out}} e^{i\omega u} + A_{\text{in}} e^{i\omega v} \tag{4}$$

where $u \equiv t - r_*$ and $v \equiv t + r_*$ are null coordinates. Expressing Eq. 4 in terms of the components of $\omega$,

$$\Psi(t, r_*) \approx A_{\text{out}} e^{-\omega_{\text{I}} u} e^{i\omega_{\text{R}} u} + A_{\text{in}} e^{-\omega_{\text{I}} v} e^{i\omega_{\text{R}} v} \tag{5}$$

The outgoing term decays exponentially with increasing $u$, as indicated by the thickness of the blue arrows in Fig. 1, while the ingoing term decays exponentially with increasing $v$, as indicated by the thickness of the red arrows. On future null infinity $\mathcal{I}^+$, we see that the outgoing term is finite, as indicated by the various blue arrows between 0 and ∞, while the incoming term is 0, as indicated by the red arrow. Thus, the wave is purely outgoing for any frequency. On the future event horizon $\mathcal{H}^+$, the red incoming arrows are finite, while the blue outgoing arrow is 0, so the wave is purely incoming for any frequency. On past null infinity $\mathcal{I}^-$, the outgoing arrow is infinite while the incoming arrows are finite, so the outgoing wave is infinitely larger than the incoming wave. On the past event horizon $\mathcal{H}^-$, the incoming arrow is infinite while the outgoing arrows are finite. However, the outgoing waves at $\mathcal{H}^-$ are unphysical because they are not smooth, so they do not exist [8], but even if they did exist, the incoming wave would be infinitely larger than the outgoing wave. Table I also expresses the relative magnitudes of the waves. The $u$ and $v$ columns of the table indicate the null coordinates on each boundary. By comparing the $e^{-\omega_{\text{I}} u}$ and $e^{-\omega_{\text{I}} v}$ columns, one can determine whether the incoming or outgoing wave is infinitely larger than the other, according to Eq. 6. The result is shown in the last column – For all boundaries with $r_* \to +\infty$, the outgoing wave is infinitely larger, while for the boundaries with $r_* \to -\infty$, the incoming wave is infinitely larger. On a side note, the boundaries featuring waves of infinite amplitude would not be relevant for physical observations, despite the fact that the boundary conditions are met [9, 10].

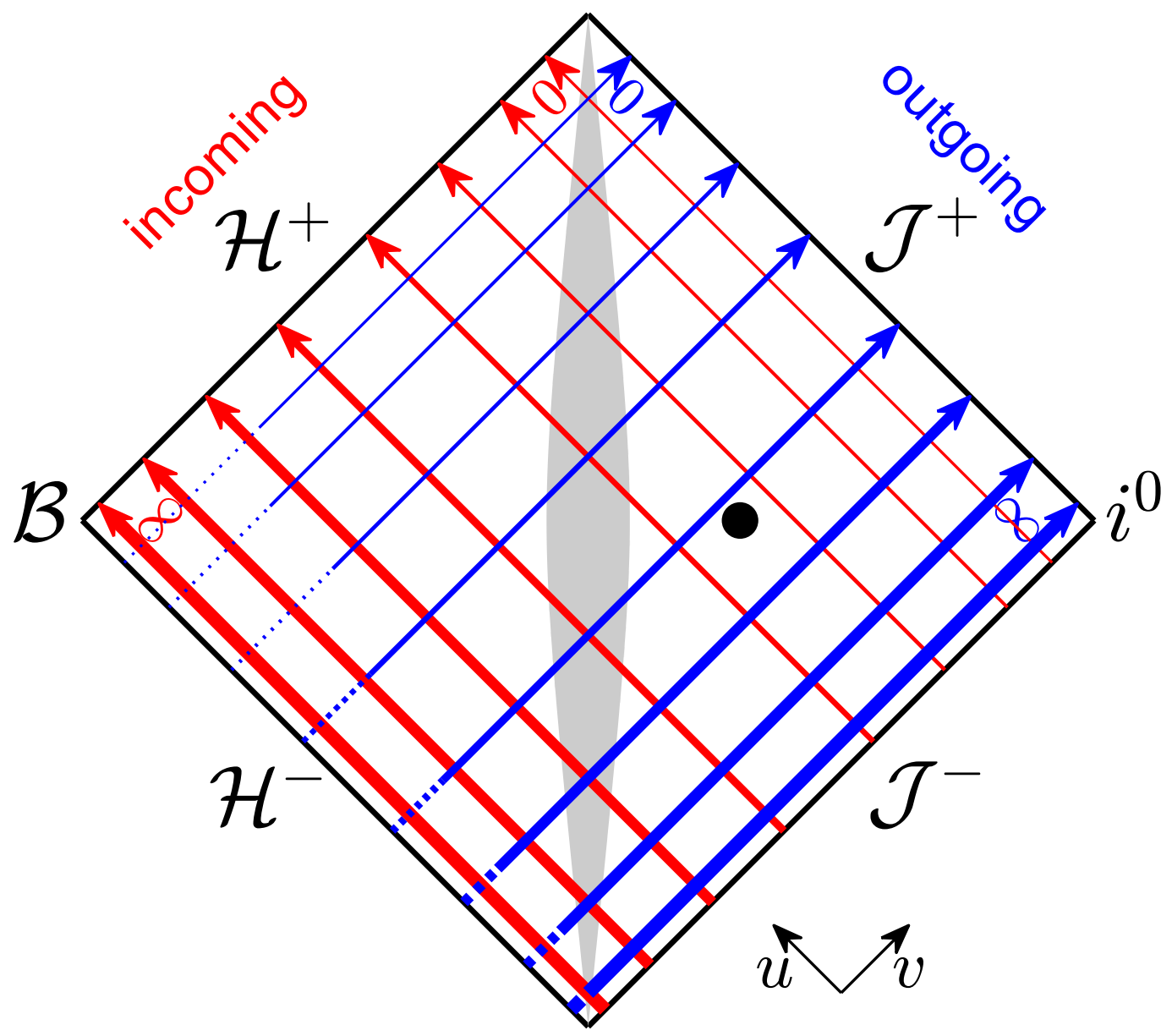

FIG. 1. Penrose-Carter conformal diagram of the domain of outer communications in the Schwarzschild spacetime. The blue (red) arrows indicate the world lines of maxima of an outgoing (incoming) wave which decays with time ($\omega_{\text{I}} > 0$). Increasing arrow thickness indicates increasing amplitude, ranging from 0 to $\infty$. The dotted blue segments indicate unphysical waves. The directions of increasing $u$ and $v$ are indicated, although the axes of the diagram are not linear in $u$ and $v$. The region of non-negligible $V$ is indicated in gray. Scattering from this region is not illustrated. The black point lies in a region of negligible $V$, but at a finite value of $r_*$.

| boundary | $u$ | $v$ | $r_*$ | $t$ | $e^{-\omega_{\text{I}} u}$ (outgoing) | $e^{-\omega_{\text{I}} v}$ (incoming) | result |
|---|---|---|---|---|---|---|---|
| $\mathcal{J}^+$ | finite const | $\to \infty$ | $\to \infty$ | $\to \infty$ | finite const | $\to 0$ | outgoing |
| $\mathcal{J}^-$ | $\to -\infty$ | finite const | $\to \infty$ | $\to -\infty$ | $\to \infty$ | finite const | outgoing |
| $i^0$ | $\to -\infty$ | $\to \infty$ | $\to \infty$ | finite const | $\to \infty$ | $\to 0$ | outgoing |
| $\mathcal{H}^+$ | $\to \infty$ | finite const | $\to -\infty$ | $\to \infty$ | $\to 0$ | finite const | incoming |
| $\mathcal{H}^-$ | finite const | $\to -\infty$ | $\to -\infty$ | $\to -\infty$ | finite const (unphysical) | $\to \infty$ | incoming |
| $\mathcal{B}$ | $\to \infty$ | $\to -\infty$ | $\to -\infty$ | finite const | $\to 0$ | $\to \infty$ | incoming |

Table 1. Evaluating the outgoing/incoming property at the boundaries of the domain of outer communications. The result column indicates the infinitely larger wave. The result is outgoing whenever $r_* \to \infty$, and incoming whenever $r_* \to -\infty$ (the horizon). This result only relies on the fact that $\omega_{\text{I}} > 0$.

The QNM boundary conditions for $\psi$ are sometimes written in the form

$$\psi \to e^{-i\omega r_*} \qquad (r_* \to \infty) \tag{6a}$$

$$\psi \to e^{i\omega r_*} \qquad (r_* \to -\infty) \tag{6b}$$

We find this statement to be ambiguous. One possible interpretation would be

$$A_{\text{out}} e^{-i\omega r_*} \gg A_{\text{in}} e^{i\omega r_*} \qquad (r_* \to \infty) \tag{7a}$$

$$A_{\text{out}} e^{-i\omega r_*} \ll A_{\text{in}} e^{i\omega r_*} \qquad (r_* \to -\infty) \tag{7b}$$

We have seen that Eq. 7 is met for any frequency with $\omega_{\text{I}} > 0$, so this interpretation of Eq. 6 does not result in a discrete spectrum of QNMs. In order to arrive at a correct interpretation, we should consider the relevant physical phenomena; QNMs should be generated by the black hole, rather than by the scattering of an incoming wave. For scattering, an incoming wave would enter at past null infinity $\mathcal{I}^-$, would scatter off the region of finite $V$ indicated in gray in Fig. 1, and would affect the amplitudes of the outgoing and incoming waves appearing at future null infinity $\mathcal{I}^+$ and the future event horizon $\mathcal{H}^+$, respectively. For a QNM, we should ensure that there is no such incoming wave at $\mathcal{I}^-$. We have shown in Fig. 1 and Table 1 that the finite incoming wave at $\mathcal{I}^-$ is dominated by an infinite outgoing wave at the same point in spacetime. Nevertheless, we must nullify the relatively minuscule incoming wave to prevent scattering. Thus, we must require [8-12]

$$A_{\text{in}} \to 0 \qquad (r_* \to \infty) \tag{8a}$$

This is the relevant interpretation of Eq. 6a. Similarly, one might consider a finite outgoing wave just outside the past event horizon $\mathcal{H}^-$, in the presence of the infinite incoming wave there. Such an outgoing wave would scatter off the region of finite $V$, and affect the amplitudes of the outgoing and incoming waves at $\mathcal{I}^+$ and $\mathcal{H}^+$. Thus, we must require

$$A_{\text{out}} \to 0 \qquad (r_* \to -\infty) \tag{8b}$$

which is the relevant interpretation of Eq. 6b. Another justification for Eq. 8b is that the outgoing wave near $\mathcal{H}^-$ is unphysical [8]. Eq. 8 serves to nullify the waves directed at the region of finite $V$, despite the presence of infinite counterpropagating waves at the same boundaries of the domain of outer communications. Eq. 8 results in a discrete spectrum of QNMs. It is equivalent to asserting that QNMs correspond to poles in the Green's function [8, 13, 14].

We can now consider finding QNMs directly from the boundary conditions. We employ the following procedure:

(i) Choose an arbitrary frequency $\omega$ in the upper half of the complex plane.

(ii) Find a solution to the wave equation (1) using this frequency, which perfectly meets the incoming boundary condition near the horizon.

(iii) Check the extent that the resulting $\psi$ obeys the outgoing boundary condition.

But which outgoing boundary condition should we use? Eq. 7a would not be applicable, since the resulting spectrum would not be discrete. Eq. 8a would be too difficult, since it would involve evaluating the amplitude of the finite incoming wave at $\mathcal{I}^-$, in the presence of an infinite outgoing wave. We thus arrive at an approximate compromise; we will consider a point in spacetime which has negligible $V$ but finite $r_*$, such as the black point in Fig. 1. This point is located to the right of the peak in $V$ shown in Fig. 2(a). Eq. 2 applies at such a point, and we will demand there

$$A_{\text{out}}e^{-i\omega r_*} \gg A_{\text{in}}e^{i\omega r_*} \qquad \text{(some finite } r_*\text{)} \tag{9}$$

Unlike Eq. 7a, this will not be met for all frequencies since the factors $e^{\pm\omega_I r_*}$ have not yet reached zero or infinity. This strategy seems reasonable since we can think of the finite localized potential as being the source of the radiation, and it would be difficult for the source to produce radiation incoming to itself. Thus, for a QNM, we expect a minimal quantity of incoming radiation even at finite distances from the potential, and not only at infinity.

Eq. 1 can be written as two coupled first order equations for the variables $\psi$ and $\psi'$,

$$\frac{\partial\psi}{\partial r_*} = \psi' \tag{10a}$$

$$\frac{\partial\psi'}{\partial r_*} = (V - \omega^2)\psi \tag{10b}$$

For a given $\omega$, if we choose finite values for $\psi$ and $\psi'$ at any finite point $r_* = a$ (near the maximum of $V$, for example), we can propagate $\psi$ and $\psi'$ via Eq. 10 to all other values of $r_*$. If we choose $\psi = 1$ and $\psi' = 0$ at point $a$, then we obtain a solution $\psi_{\text{even}}$ which approximates an even function about $a$. If we choose $\psi = 0$ and $\psi' = 1$ at point $a$, then we obtain a solution $\psi_{\text{odd}}$ which approximates an odd function about $a$. These two solutions are necessarily independent since their behaviors at point $a$ are fundamentally different. Since Eq. 1 is linear, we can form a solution

$$\psi = A\psi_{\text{even}} + B\psi_{\text{odd}} \tag{11}$$

for any finite, complex values $A$ and $B$. This solution has $\psi = A$ and $\psi' = B$ at point $a$, so it includes all possible values of $\psi$ and $\psi'$ at point $a$. Thus, this is a general solution. In other words, each frequency is doubly degenerate; there are exactly two independent wavefunctions for each complex frequency.

For the sake of evaluating the quantity of incoming radiation, we would like a method of determining if a given solution is incoming or outgoing at a given point in space. If a wave is incoming in a region of negligible $V$, then it is of the form $\psi \propto \exp(i\omega r_*)$. On the other hand, if $V$ is non-zero but slowly varying, we can approximate that $\psi \propto \exp(i\omega_{\text{eff}} r_*)$, where $\omega_{\text{eff}}(r_*) = \sqrt{\omega^2 - V(r_*)}$. This implies that $\partial\psi/\partial r_* = i\omega_{\text{eff}}\psi$ for an incoming wave. An outgoing wave obeys $\partial\psi/\partial r_* = -i\omega_{\text{eff}}\psi$. Thus, we can quantify the extent that a wave is incoming or outgoing via the quantity

$$\phi = \frac{1}{i\omega_{\text{eff}}\psi}\frac{\partial\psi}{\partial r_*} \tag{12}$$

This quantity is unity for a purely incoming wave, and -1 for a purely outgoing wave. Any other complex value implies a superposition of incoming and outgoing waves. Comparing $\phi$ with $\pm 1$ and averaging over a finite spatial window specified by the function $f(r_*)$ gives a measure of the purity

$$P \equiv -\frac{1}{2}\ln\int_{-\infty}^{\infty} dx\, f(r_*)\, |\phi \pm 1|^2 \tag{13}$$

where the top (bottom) sign corresponds to outgoing (incoming) waves. Larger values of $P$ correspond to fewer waves incoming toward the peak in the effective potential, which implies a stronger QNM. The use of the natural logarithm is merely for the sake of visualization via plots.

For illustration, we consider the fundamental, least-damped QNM for $l = 2$, which has $\omega_R = 0.747343$ and $\omega_I = 0.177925$ in units of $1/2M$, where $M$ is the mass of the black hole [4, 5]. We choose the point $a$ to be $r_* = 1.066$ in units of $M$, which is close to the maximum of the Zerilli potential shown in Fig. 2(a) [7]. Starting with $\psi = 1$ and $\psi' = 0$ at $a$, we integrate Eq. 10 numerically to the right and left. The resulting $\psi_{\text{even}}$ is shown in Fig. 2(b) [15]. Starting with $\psi = 0$ and $\psi' = 1$, we obtain $\psi_{\text{odd}}$, as shown in Fig. 2(c).

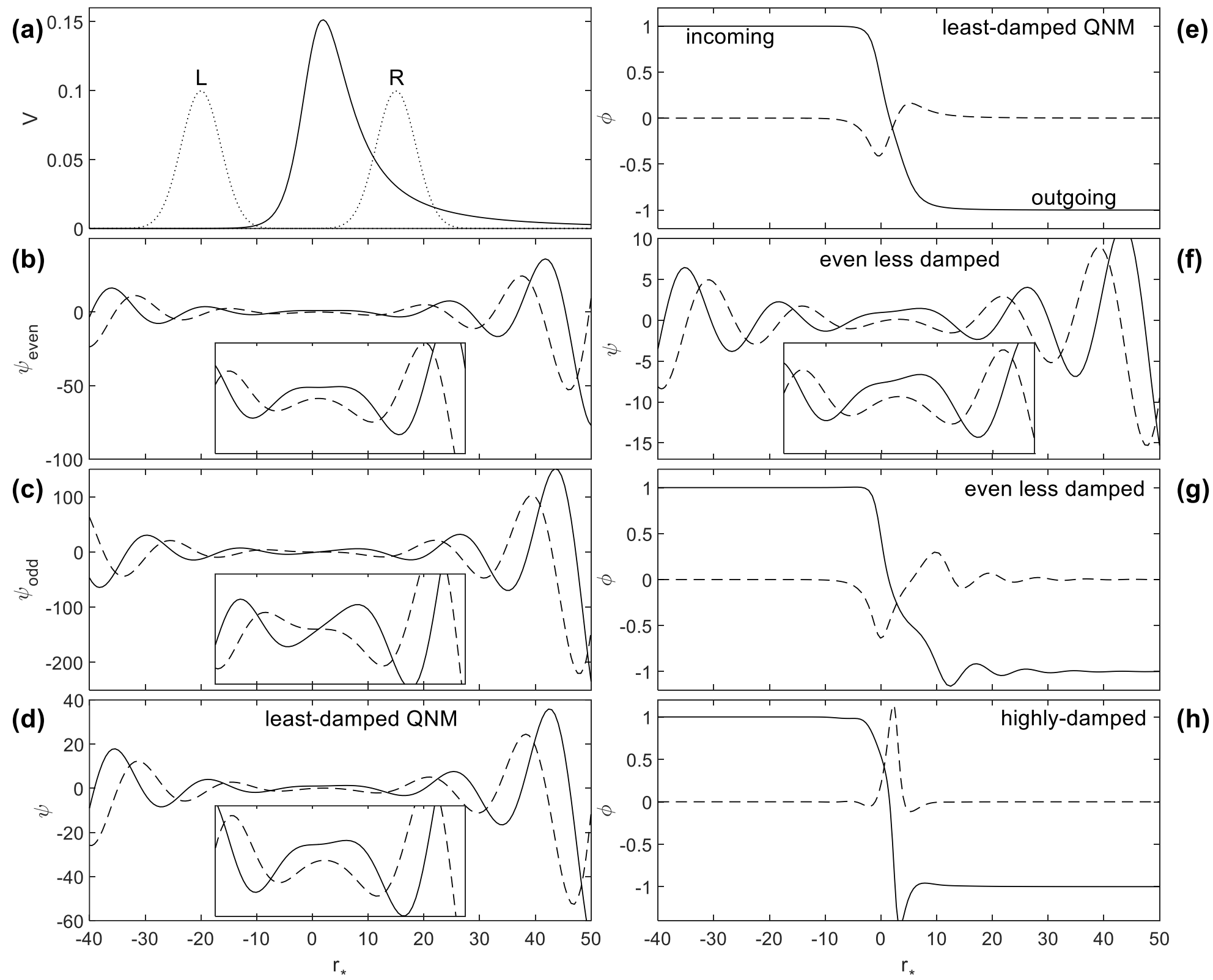

FIG. 2. Pure and impure modes with $l = 2$ in a Schwarzschild black hole. Solid (dashed) curves indicate real (imaginary) parts. The insets show enlargements of the central regions. (a) The Zerilli potential. The dotted curves indicate the left window $L$ used to find the optimal QNM, and the right window $R$ used to analyze its purity. (b) and (c) Even and odd solutions respectively, for the least-damped QNM. (d) The optimal wavefunction for the least-damped QNM. (e) $\phi$ for the wavefunction shown in (d). (f) and (g) Like (d) and (e), but for a frequency with less damping than the least-damped QNM. (h) $\phi$ for a frequency with damping as high as the first overtone, but a different value of $\omega_R$.

We would like to find the superposition of $\psi_{\text{even}}$ and $\psi_{\text{odd}}$ which is purely incoming in the region to the left of the peak in the effective potential. Since Eq. 1 is linear, overall constants in Eq. 11 are not important. Thus, we set $A = 1$, and search in the complex plane for the value of $B$ which maximizes $P$ within the left gaussian window, using the bottom sign in Eq. 13 since we are interested in incoming waves. The resulting wavefunction is shown in Fig. 2(d). For this particular QNM, the optimal $B$ is small, so $\psi$ is similar to $\psi_{\mathrm{even}}$. Fig 2(e) shows $\phi$ of Eq. 12, computed from $\psi$ of Fig. 2(d). The unity value in the left region indicates the incoming wave which we found by optimizing $B$. The value -1 obtained in the right region reflects the purity of the least-damped QNM due to its lack of an incoming wave, i.e., it is almost purely outgoing. In contrast, Fig. 2(g) shows $\phi$ for a mode with 0.7 of the damping of the least-damped QNM (but the same $\omega_{\mathrm{R}}$). In this case, we see that the right region has an oscillation. Since $\phi$ is not -1, the mode has a significant incoming component.

In principle, we could have found the optimal $\psi$ for the given $\omega$ by choosing an arbitrary value for $\psi$ at some point well to the left of the potential peak, and also choosing $\partial\psi/\partial r_*$ at that point to be given by $i\omega\psi$, which corresponds to a purely incoming solution. We could have numerically integrated Eq. 10 to the right (for increasing $r_*$). However, this involves propagating the solution inward toward the potential peak, a process which is plagued with numerical errors, as described in Ref. 4. Thus, we employ the strategy discussed above using $\psi_{\mathrm{even}}$ and $\psi_{\mathrm{odd}}$, which only involves outward integration.

We quantify the purity for each trial frequency by computing $P$ within the right gaussian window of Fig. 2(a), using the top sign in Eq. 13. The location of this window (centered at 15 with a gaussian width $\sigma = 5$) is schematically represented by the black point in Fig. 1, and constitutes the finite value of $r_*$ in Eq. 9. The smooth gaussian window prevents the appearance of spurious "pure" modes. Fig. 3(a) shows the result for the $l = 2$ gravitational mode spectrum. Lighter gray indicates purer modes. One outstanding mode appears as a white spot near the lower edge of the figure. This spot agrees well (within 1%) with the fundamental least-damped QNM, indicated by a dashed blue circle and by dotted lines in the horizontal and vertical profiles. On the other hand, the solid blue circle indicates the position of the first overtone, but no white spot is seen there. We can understand this by inspecting Fig. 2(h), which is computed with the same $\omega_{\mathrm{I}}$ as the first overtone, but $\omega_{\mathrm{R}}$ 1.2 times larger. The solid curve in Fig. 2(h) goes rapidly to -1 since the incoming component decays to zero over a very short distance from the potential peak. Thus, all the highly-damped modes have high values of $P$, and the first overtone is no purer than the neighboring modes, within the precision of the analysis. The value of $P$ within the solid blue circle is even larger than that of the least-damped QNM, due to the trend that $P$ increases with $\omega_{\mathrm{I}}$, as seen in the vertical profile of Fig. 3(a). Similar results are obtained for the $l = 3$ gravitational and $l = 1$ electromagnetic modes shown in Figs. 3(b) and 3(c). In both cases, the least-damped QNM (dashed blue circle) corresponds to a strong white spot, while no special significance is seen for the first overtone (solid blue circle). In the electromagnetic case, the QNM frequencies were taken from Refs. 3 and 16. Fig. 3(d) shows the purity of neutrino modes (neglecting the rest mass), computed with the effective potential given in Refs. 6 and 17 for angular momentum parameter $\kappa = 1$. The single white spot in Fig. 3(d) seems to be the fundamental QNM, although we do not find a prediction in the literature for comparison.

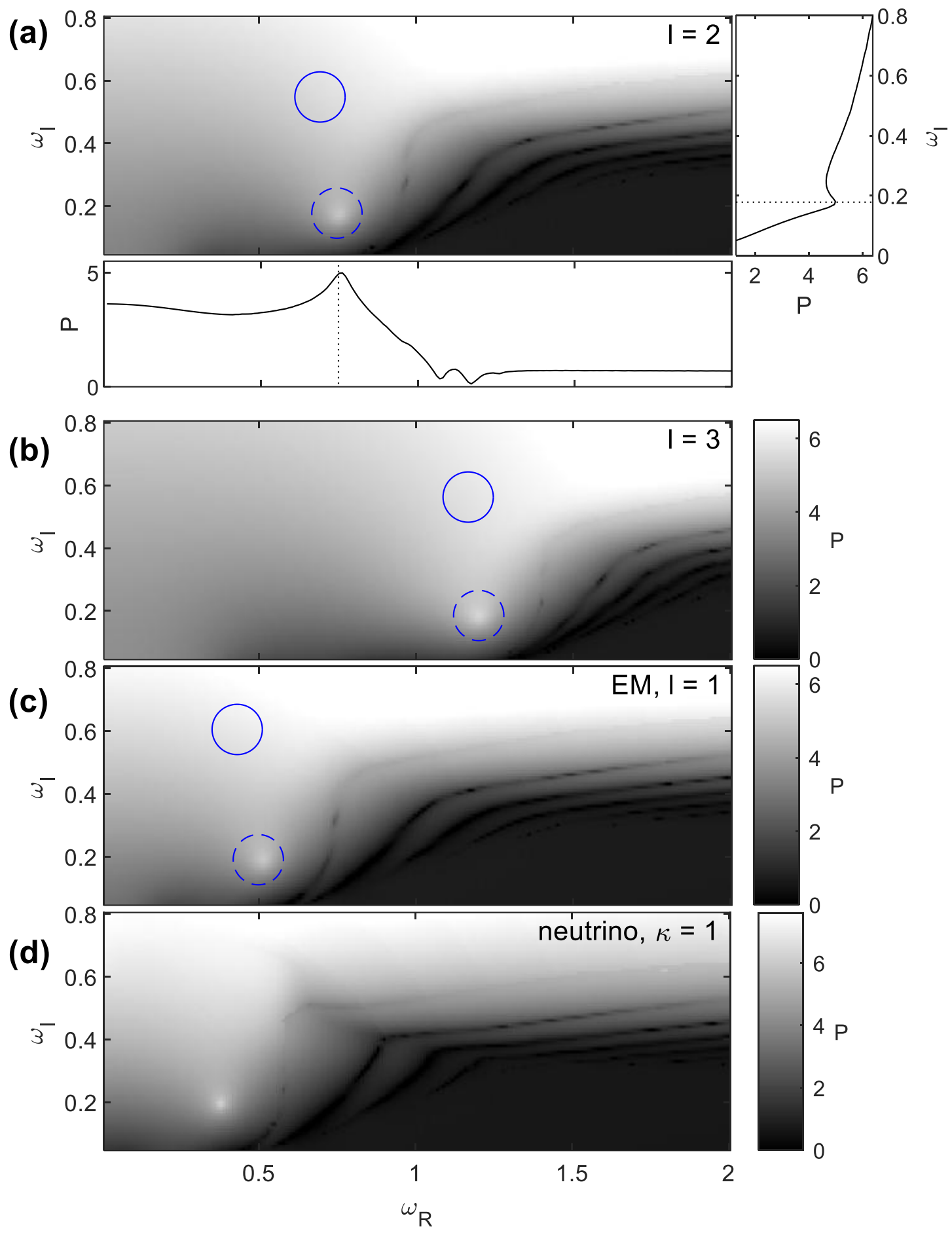

FIG. 3. Searching for QNMs. The purity $P$ of the modes is shown. Dashed blue circles and dotted lines indicate the fundamental least-damped QNM. Solid blue circles indicate the first overtone. (a) The $l = 2$ gravitational modes for the Zerilli potential. The profiles are cuts through the local maximum at the white spot. (b) The $l = 3$ gravitational modes. (c) The $l = 1$ electromagnetic modes. (d) The $\kappa = 1$ neutrino modes.

The panels of Fig. 3 also display a black region of especially low purity near the lower right corner. In this region, $\omega_R$ is so large that $V$ is negligible. In this case, the incoming solution which we choose to the left of the potential peak continues across the peak to the right.

The results shown in Fig. 3(a) depend weakly upon the choice of the right gaussian window, although the QNM peak is robust. If the center of the window is moved from 15 to 25, the visibility of the fundamental mode decreases, as seen in Fig. 4(a). If the center is moved to 10, the white spot corresponding to the fundamental mode broadens and becomes less visible as shown in Fig. 4(b); apparently $\exp(-i\omega_{\text{eff}}r_*)$ is not a good approximation for the wavefunction so close to the maximum in the effective potential. If the gaussian width of the window is decreased from 5 to 2, the window becomes too sharp, which causes the appearance of spurious white peaks, as seen in Fig. 4(c). These peaks become even more clear if the width of the window is further decreased to unity, as seen in Fig. 4(d).

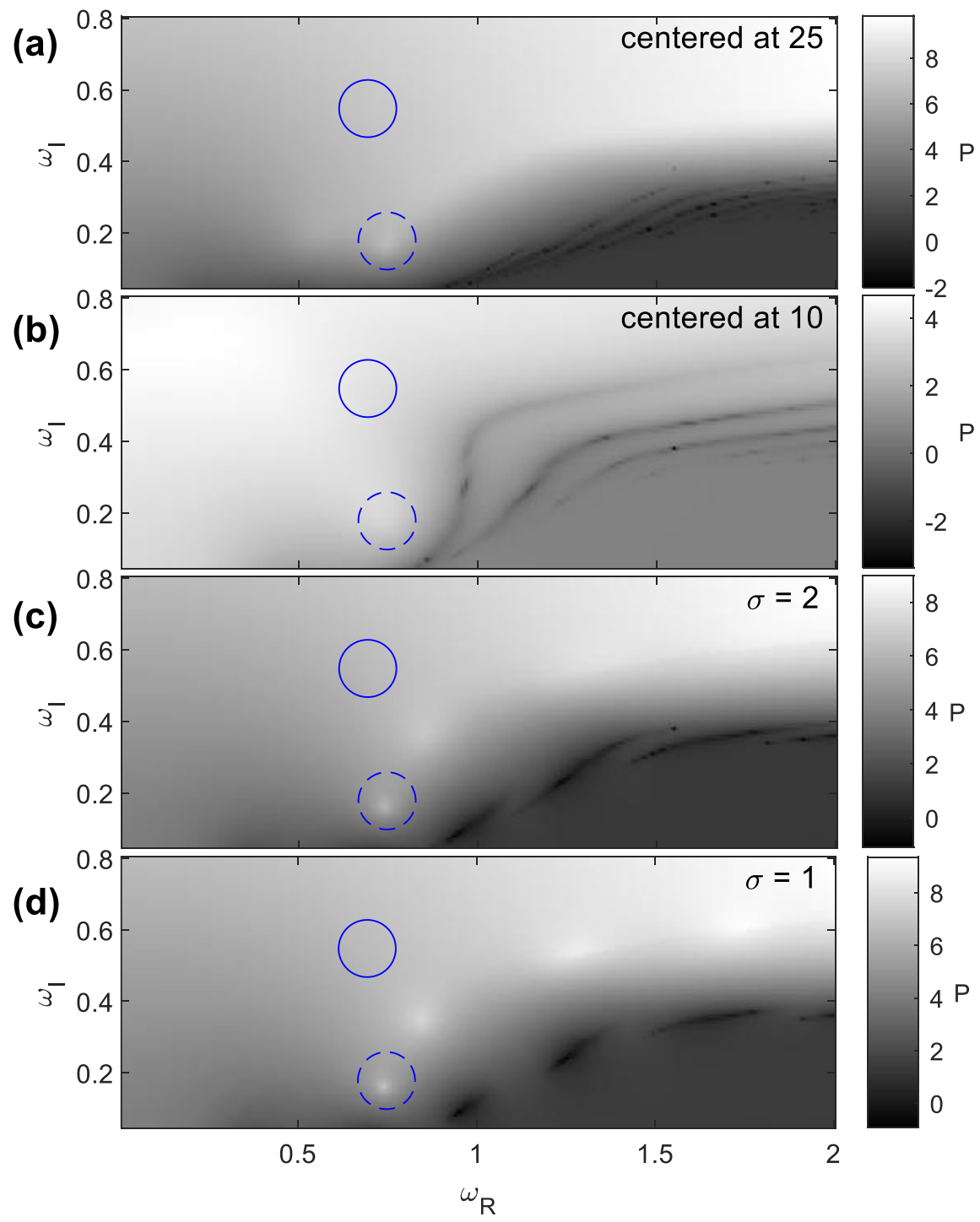

FIG. 4. The effect of modifying the right window. The purity $P$ of the modes is shown. Dashed blue circles and dotted lines indicate the fundamental least-damped QNM. Solid blue circles indicate the first overtone. The right window is modified to be (a) centered at 25. (b) centered at 10. (c) of width $\sigma = 2$. (d) of width $\sigma = 1$.

In conclusion, we find the fundamental QNMs by directly looking for wavefunctions which satisfy the incoming/outgoing boundary conditions. This is conceptually satisfying and simple to implement, although the technique is limited to the case of low damping. For higher damping, any incoming mode is overpowered by the outgoing mode even at finite distances, so any frequency seems to meet the boundary conditions. Thus, even the first overtone is out of our reach. Also, the method is inherently approximate since the boundary condition is applied at a finite distance, which results in limited accuracy, in contrast to other techniques. Furthermore, we see that one must be careful when writing the QNM boundary conditions; the outgoing condition at infinity implies the nullification of the incoming mode at past null infinity, despite the presence of an infinite outgoing mode there.

**Acknowledgments:** I thank A. Ori, S. Hadar, V. Cardoso, K. Kokkotas, K. Destounis, D. Hilditch, V. Boyanov, J. E. Patiño, J. D. Álvares and J. Santos for helpful discussions. This work was supported by the Israel Science Foundation, grant 531/22.